\documentclass[5p,twocolumn,preprint,longtitle]{elsarticle}
\usepackage[pagewise]{lineno}
\usepackage{amsmath,amsfonts,amssymb,amsthm}
\usepackage{cancel}
\usepackage{appendix}
\usepackage{fancyhdr}   
\usepackage{url}
\usepackage{path}
\usepackage{tabularx}
\usepackage{array,hhline,dcolumn}

\begin{document}
\title{First Measurement of $\theta_{13}$ From Delayed Neutron Capture on Hydrogen in the Double Chooz Experiment
}

\author[TokyoInst]{Y.~Abe}

\author[MaxPlanck]{C.~Aberle}

\author[CBPF]{J.C.~dos Anjos}

\author[CEA]{J.C.~Barriere}

\author[Davis]{M.~Bergevin}

\author[Livermore]{A.~Bernstein}

\author[TohokuUni]{T.J.C.~Bezerra}

\author[INR]{L.~Bezrukhov}

\author[Chicago]{E.~Blucher}

\author[Livermore]{N.S.~Bowden}


% ...
\address[CBPF]{Centro Brasileiro de Pesquisas F\'{i}sicas, Rio de 
Janeiro, RJ, cep 22290-180, Brazil}
\address[Chicago]{The Enrico Fermi Institute, The University of 
Chicago, Chicago, IL 60637, USA}
% ...
\address[Davis]{University of California, Davis, CA-95616-8677, USA}
% ...
\address[INR]{Institute of Nuclear Research of the Russian Aacademy 
of Science, Russia}
\address[CEA]{Commissariat \`{a} l'Energie Atomique et aux Energies 
Alternatives, Centre de Saclay, IRFU, 91191 Gif-sur-Yvette, France}
\address[Livermore]{Lawrence Livermore National Laboratory, 
Livermore, CA 94550, USA}
% ...
\address[MaxPlanck]{Max-Planck-Institut f\"{u}r Kernphysik, 69117 
Heidelberg, Germany}
% ...
\address[TohokuUni]{Research Center for Neutrino Science, Tohoku 
University, Sendai 980-8578, Japan}
% ...
\address[TokyoInst]{Department of Physics, Tokyo Institute of 
Technology, Tokyo, 152-8551, Japan  }
% ...

\begin{abstract}
The Double Chooz experiment has determined the value of the neutrino oscillation parameter $\theta_{13}$\  from an analysis of inverse beta decay interactions with neutron capture on hydrogen. This analysis uses a three times larger fiducial volume than the standard Double Chooz assessment, which is restricted to a region doped with gadolinium (Gd), yielding an exposure of 113.1 GW-ton-years. 
The data sample used in this analysis is distinct from that of the Gd analysis, and the systematic uncertainties are also largely independent, with some exceptions, such  as
 the reactor neutrino flux prediction. 
A combined rate- and energy-dependent fit finds $\sin^2\!2\theta_{13}=0.097\,\pm\,0.034\,\rm{(stat.)} \,\pm\,0.034\,\rm{(syst.)}$, excluding the no-oscillation hypothesis at $2.0\,\sigma$. 
This result is consistent with previous measurements of $\sin^2\!2\theta_{13}$. 
\end{abstract}

\maketitle

%
%
Neutrino oscillations are well established in the three flavor paradigm and can be described by three mixing angles ($\theta_{12}$, $\theta_{23}$, $\theta_{13}$), a CP-violating phase $\delta$,  and two mass-squared differences ($\Delta m_{21}^2, \Delta m_{32}^2$). Among the three mixing angles, 
 $\theta_{13}$ is the smallest and has recently been revealed to be non-zero~\cite{T2KTheta13,MINOSTheta13,DC1stPub,DB1stPub,RENO1stPub,DC2ndPub,DB2ndPub}. 
The value of $\theta_{13}$\ is a critical input for plans to measure $\delta$\ and the neutrino mass hierarchy. 
Furthermore, it may provide important clues for physics beyond the Standard Model. 
The current  best measurements of $\theta_{13}$~  come from the reactor $\bar{\nu}_e$-disappearance experiments Double Chooz, Daya Bay, and RENO~\cite{DC2ndPub,DB2ndPub,RENO1stPub}. 
All three experiments rely on the detection of the inverse beta decay (IBD) interaction, $\bar{\nu}_e+p\rightarrow e^++n$, in  Gd-doped liquid scintillator (LS). 
Typically these experiments search for a prompt positron signal followed by an $\sim8$~MeV gamma cascade from neutron capture on Gd. Background due to  natural radioactivity, which is predominantly below 4~MeV, is largely suppressed.
However, in Double Chooz it is also possible to search for a prompt positron followed by a 2.2~MeV gamma ray from neutron capture on hydrogen, thanks to the low background environment in the detector. 

Though the latter analysis presents several challenges, it provides important benefits: a cross-check on the standard Gd analysis and  improved $\bar{\nu}_e$ energy spectrum shape information which is essential to our knowledge of $\theta_{13}$.

In this letter we present an analysis of IBD interactions with neutron capture on hydrogen in the Double Chooz far detector. 
Following the same approach as in  previous reports~\cite{DC1stPub,DC2ndPub}, this analysis compares the candidate event rate and prompt energy spectrum shape to the Monte Carlo (MC) prediction. 
This analysis, however, differs from those reported~\cite{DC1stPub,DC2ndPub} in two major ways. 
First, the definition of the delayed signal is changed from the $\sim8$~MeV gamma cascade characteristic of a neutron capture on Gd to the 2.2~MeV gamma ray characteristic of a neutron capture on hydrogen. 
This change allows us to select a data set that is statistically independent of the Gd-based data set and has different systematic uncertainties and background characteristics. 
Second, because hydrogen captures occur in the undoped LS in addition to the Gd-doped region, a three times larger fiducial volume is available for analysis.  

The Double Chooz far detector is located at a distance of $\sim1050$~m from the two $4.25\,\rm GW_{th}$\ reactor cores of the Chooz Nuclear Power Plant, with a rock overburden of $ 300$~meters  water equivalent.
The central region of the detector consists of three concentric cylinders, collectively called the inner detector (ID). The innermost cylinder is the 10.3~m$^3$\ target. This is surrounded by a $\gamma$-catcher (22.5~m$^3$).  
The target liquid is a PXE-based LS doped with Gd at a concentration of 1~g/l~\cite{scintillator}, while the $\gamma$-catcher liquid is an undoped LS. 
Outside the $\gamma$-catcher is the buffer, a 105~cm thick layer of non-scintillating mineral oil contained in a stainless steel tank. Light from the target and $\gamma$-catcher volumes is collected by 390 low-background 10-inch PMTs installed on the inner wall of the buffer tank~\cite{PMT1,PMT2,PMT3}. 
Outside the buffer tank, and optically isolated from it, is 
the inner veto (IV), a 50~cm thick layer of liquid scintillator in a steel tank. 
The IV is equipped with 78 8-inch PMTs and serves as a veto for cosmic rays and fast neutrons entering the detector. The IV is surrounded by a 15~cm thick layer of demagnetized steel which  suppresses $\gamma$-rays from radioactivity in the surrounding rock. 
Above the IV is the outer veto (OV) detector, a scintillator-strip-based muon tracking system. 
The OV system was installed during the data taking period, and about 2/3 of the data in this analysis benefit from OV use. 
A more detailed description of the entire detector can be found in Ref.~\cite{DC2ndPub}.

The number of protons  is estimated to be $(6.747\,\pm\,0.020)\times10^{29}$\  in the target~\cite{DC2ndPub} and $(1.582\,\pm\,0.016)\times10^{30}$\  in the $\gamma$-catcher volume, the latter being based on a geometrical survey and measurements of the scintillator hydrogen fraction.

The IBD signal is a twofold coincidence of a prompt positron energy deposition, $\rm E_{prompt}$, and a delayed gamma energy deposition, $\rm E_{delay}$, resulting from a neutron capture on hydrogen or Gd.  
The separation in time and space, $\Delta$t and $\Delta$r,  of the coincident events are determined by neutron capture physics. 
Neutron capture times are 200~$\mu$s in the  $\gamma$-catcher and 30~$\mu$s in the target, where the presence of Gd greatly increases the neutron capture probability.
In this analysis, where we search for $\rm E_{delay}\approx2.2$~MeV without any fiducial volume cuts, we expect to detect candidates in both the target and $\gamma$-catcher.
Given that only 13~\% of the IBD interactions in the target volume are followed by neutron capture on hydrogen~\cite{DC2ndPub}, 95~\% of the signal events used in this analysis are located in the  $\gamma$-catcher.

Vertex reconstruction is based on a likelihood maximization of the charge and timing of the pulses detected at each PMT~\cite{DC2ndPub}. 
It allows the spatial correlation of prompt and delayed events, effectively removing accidental backgrounds.

We reconstruct the energy of all events via two steps: (1) a total charge ($\rm Q_{tot}$) to photoelectron ($\rm PE_{tot}$) conversion; and (2) a $\rm PE_{tot}$\  to visible energy ($\rm E_{vis}$) conversion as done in the Gd analysis~\cite{DC2ndPub}. 
The first step takes into account a channel-by-channel, non-linear gain calibration. 
The second step uses a light yield of $\sim230$~PE/MeV, defined by the neutron capture peak on hydrogen in $^{252}$Cf calibration source data.   
By applying correction factors derived from spallation neutron data, this step also corrects for the time variation and vertex dependence of the detector response.
The same method is used to determine $\rm E_{vis}$\  for the MC sample.

This analysis uses data collected by the Double Chooz far detector between April 13, 2011 and March 15, 2012, which is the same time-period used in the latest Double Chooz Gd analysis~\cite{DC2ndPub}. 
The total live time is  240.1 days, which is different from 227.9 days used in the Gd analysis~\cite{DC2ndPub} because of different analysis cuts. 

%
%
The IBD candidate selection is performed via the following procedure. 
To reduce muon-induced backgrounds, we reject all events that occur less than 1~ms after a cosmic muon crosses the IV or the ID. 
We use PMT charge isotropy and PMT pulse simultaneity cuts  to reduce backgrounds caused by light emitted from PMT bases (``light noise'')~\cite{DC1stPub}.
We apply the following coincident selection cuts to the remaining events: $\rm 0.7~MeV< E_{prompt}<12.2~MeV$; $\rm 1.5~MeV<E_{delay}<3.0~MeV$; 
$\rm 10~\mu s<\Delta t<600~\mu s$; $\rm \Delta r<90~cm$. 
Furthermore, we reject prompt candidates that are  coincident with a signal  detected in the OV. This veto, along with the 10~$\mu$s lower bound of the $\Delta$t cut, renders backgrounds due to stopped muons negligible.
Finally, we apply a multiplicity cut to reduce fast neutron backgrounds. 
This cut demands that no trigger occur in the 600~$\mu$s preceding the prompt candidate and that no trigger other than the delayed candidate occur in the 1000~$\mu$s following the prompt candidate. 

The selection cuts yield a total of $36284$ events. 
Among these IBD candidates are backgrounds due to uncorrelated accidental coincidences, fast neutrons produced by muons traversing the nearby rock,  long-lived cosmogenic isotopes (mainly $^{9}$Li), and a small contribution from light noise.  
Accidentals are the dominant background, comprising almost half the IBD candidate sample.

%
%
We measure the rate and energy spectrum of accidentals by analyzing a sample of  off-time coincidences. We collect this sample by looking for a delayed trigger between $\rm 1~s+10~\mu s$ and $\rm 1~s+600~\mu s$ after a prompt candidate event and applying a multiplicity cut for a period of $\rm 1~s-600~\mu s$ to $\rm 1~s+1000~\mu s$. To increase sample statistics, we open 124 consecutive windows after this first window, thus sampling accidentals between 1~s and 1.2~s after each prompt candidate. 
After correcting for inefficiencies associated with this selection method, we obtain an accidentals rate of $73.45\,\pm\,0.16$~events/day.
The result is cross-checked among multiple independent methods, and the quoted value includes the largest systematic uncertainty among them.

%
%
The fast neutron background consists of a proton recoil, the prompt event, in coincidence with the capture of the neutron, the delayed event. 
A single muon passing close to the detector 
may generate one or more  fast neutrons which traverse the IV and ID. 
We tag the number of IBD candidates in which fast neutrons are recorded simultaneously in the IV and ID by requiring $\geq~2$ IV PMT hits and an 
ID-IV pulse-timing correlation. 
We estimate the tagging efficiency from an event sample with $\rm E_{prompt}>12~MeV$, following the same method as  used for the Gd analysis~\cite{DC2ndPub}.
From this sample we obtain a spectrum shape and, using the tagging efficiency and sample purity, we calculate the fast neutron rate to be $2.50\,\pm\,0.47$~events/day.

%
%
Muon-induced radioactive isotopes which emit a neutron immediately following $\beta$-decay, such as $^{9}$Li, can be a background to IBD reactions.
As the lifetime  of   $^9$Li is  257~ms, we use the correlation of the $^9$Li decay events  to previously detected muons to estimate the $^9$Li background rate. 
To increase the purity of $^9$Li in our sample, we consider only the subset of IBD candidates for which  the spatial separation between the prompt event and the reconstructed muon track is within a defined distance. 
While ID PMTs are used to reconstruct the muon tracks in the Gd analysis~\cite{DC2ndPub}, IV PMTs are used in this analysis to account for muons going through non-scintillating buffer liquid.
To estimate the $^9$Li rate in this subsample, we fit the time difference $\rm \Delta \rm{t}_\mu$\ between the IBD candidate prompt events and preceding muons with an exponential function characterized by the $^9$Li lifetime, plus a flat function to accommodate remaining accidentals and IBD candidates. 
The estimated rate is found to be consistent with that in the Gd analysis~\cite{DC2ndPub}, accounting for the different fiducial volumes and selection efficiencies, and the difference is included in the systematic uncertainty. 
We find a $^9$Li rate of $2.8\,\pm\,1.2$~events/day. 
Muon track reconstruction efficiency is evaluated by a MC study and added into the systematic uncertainty. 
We estimate the shape and associated systematic uncertainty from MC, as was done in the Gd analysis~\cite{DC2ndPub}.
\begin{figure}[btp]
\begin{center}
\includegraphics[width=\columnwidth]{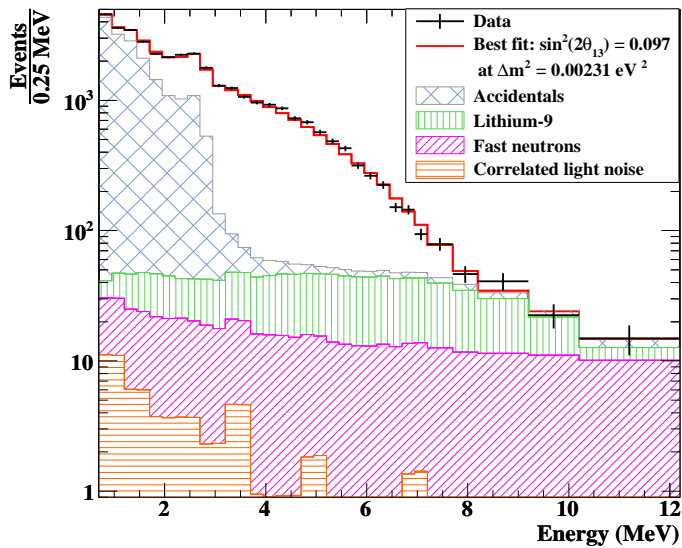}
\caption{(Color online)  Stacked histogram showing the prompt energy spectrum of neutrino candidates without background subtraction (black data points with statistical error bars). The red (grey) line is the best fit oscillation hypothesis. 
Also shown are contributions from accidentals (blue cross-hatched), $^9$Li (green vertical lines), fast neutrons (purple diagonal lines), and correlated light noise (orange horizontal lines).}
\label{fig:backgr}
\end{center}
\end{figure}

%
%
Finally, we found  a small number of light noise events creating two consecutive triggers that are  identified as IBD candidates.
A volume cut on the reconstructed vertex is used to quantify the rate and $\rm E_{prompt}$ spectrum shape for this type of background.
We estimate this background rate as $0.32\,\pm\,0.07$~events/day.

\begin{table}[tbp]
\caption{Summary of the number of observed IBD candidates and the predictions for the  
signal and background contributions  used as input for the oscillation fit analysis.}
\begin{tabular}{lr}
\hline
\hline
 \bf{Source} & \bf{Predicted/observed }\\
& \bf {events} \\
\hline
$\bar{\nu}_e$\,prediction\,(no\,osc.) & 17690\\
Accidentals   & 17630\\
Cosmogenic isotopes    & 680\\
Fast neutrons & 600\\
Light noise   & 80 \\
\hline
Total prediction & 36680 \\
\hline
Observed IBD candidates & 36284\\
\hline
\hline
\end{tabular}
\label{tab:background}
\end{table}

%
%
In this analysis, neutron detection efficiency $\rm\epsilon_n$ includes both the efficiency of the IBD selection and the fraction of neutron captures which occur on hydrogen. 
We evaluate $\rm\epsilon_n$ from $^{252}$Cf neutron source calibration data and find it to be  $\rm\epsilon_n=0.0846\,\pm\,0.0018$~in the target and $\rm\epsilon_n=0.7853\,\pm\,0.0036$~in the $\gamma$-catcher. 
Weighting by the fraction of predicted IBD candidates in each region, we estimate the uncertainty in the detection efficiency  over the entire 
fiducial volume as 1.0~\%. 
Finally, we find that an uncertainty of 1.2~\% accounts for the MC modeling of neutron migration, called spill-in/out~\cite{DC2ndPub}, between detector subvolumes.
Adding these factors in quadrature, we obtain a total detection efficiency uncertainty of 1.6~\%.

Energy scale uncertainty arises from three sources: time variation, non-linearity, and non-uniformity in the detector response. 
We treat the first two effects exactly as in Gd analysis~\cite{DC2ndPub}. 
The third effect has a larger impact on the hydrogen analysis because of its extended fiducial volume. We estimate it by comparing data and MC from calibration source deployments in the $\gamma$-catcher. 
In total, we find an energy scale uncertainty of 1.7~\%, as compared to 1.1~\% used in the Gd analysis~\cite{DC2ndPub}.

%
%
The reference $\rm E_{prompt}$~spectrum is selected from the same  reactor power-based $\bar{\nu}_e$~MC sample generated  for the Gd analysis~\cite{DC2ndPub}. 
Systematic uncertainties on the reference spectrum are the same as for the Gd analysis. 
We use the Bugey4 measurement to minimize the systematic uncertainty on the reactor neutrino flux prediction~\cite{Bugey4,DC2ndPub}, which is the dominant uncertainty in this analysis.
The  no-oscillation expectation for the number of  neutrino candidates is $36680\,\pm\,520$, including background.
The predicted number of events for both signal and backgrounds are summarized in Tab.~\ref{tab:background},
and  uncertainties relative to the predicted signal statistics are shown in  Tab.~\ref{tab:systematics}. 

\begin{figure}[tbp]
\begin{center}
\includegraphics[width=\columnwidth]{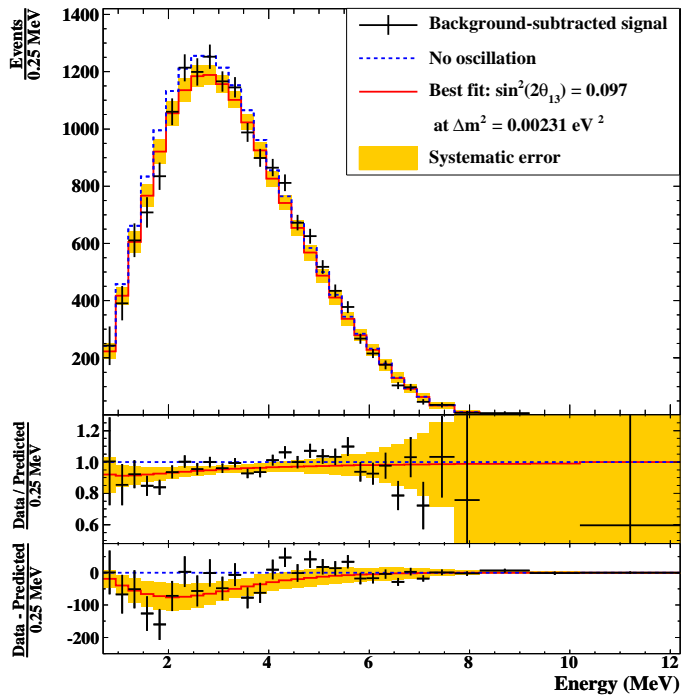}
\caption{(Color online)  Top: Background-subtracted data (black points with statistical error bars) are superimposed on the prompt energy spectra expected in the case of no oscillations (dashed blue line) and for our best fit $\sin^2\!2\theta_{13}$ (solid red line). The best fit has $\chi^2$/DOF of 38.9/30. Solid gold bands indicate systematic errors in each bin. Middle: The ratio of  data to the no-oscillation prediction (black points with statistical error bars) is superimposed on the expected ratio in the case of no oscillations (blue dashed line) and for our best fit $\sin^2\!2\theta_{13}$ (solid red line). 
Gold bands indicate systematic errors in each bin.  
Bottom: The difference between data and the no-oscillation prediction is shown in the same style as the ratio (above).}
\label{fig:ffit}
\end{center}
\end{figure}

\begin{table}[tbp]
\caption{Summary of signal and background normalization uncertainties relative to the predicted signal.}
\begin{tabular}{lr}
\hline
\hline
\textbf{Source} & \textbf{Uncertainty} [\%]\\
\hline
Reactor flux & 1.8 \\
Statistics & 1.1 \\
Accidental background & 0.2 \\
Cosmogenic isotope background & 1.6 \\
Fast neutrons & 0.6 \\
Light noise & 0.1\\
Energy scale & 0.3 \\
Efficiency & 1.6 \\
\hline
Total & 3.1\\
\hline
\hline
\end{tabular}
\label{tab:systematics}
\end{table}
%
%

To extract $\sin^2\!2\theta_{13}$\ we compare both the rate and shape of the data to the reference $\rm E_{prompt}$\  spectrum in 31 variably sized energy bins from 0.7 to 12.2~MeV. 
The fit procedure is identical to that used in the Gd analysis~\cite{DC1stPub,DC2ndPub}, except that we use a single integration period and include the $\Delta$r cut efficiency as an additional source of uncertainty. 
We use the  MINOS value of $\Delta m^2=(2.32\,\pm\,0.12)\times10^{-3}\rm{eV}^2$\  as  input for  the fit~\cite{MINOS:DM2}.
We find a best fit of 
\begin{equation*}
\sin^2\!2\theta_{13}=0.097\,\pm\,0.034\,\rm{(stat.)\,\pm\,0.034\,(syst.)}
\end{equation*}
with $\chi^2$/DOF of 38.9/30. As in the Gd analysis~\cite{DC2ndPub}, we define statistical error as the portion of the $1~\sigma$\  error which can be improved by collecting more data. This includes uncertainty from our current statistics (see Tab.~\ref{tab:systematics}) and uncertainty on background shapes. We define systematic error as the uncertainty which cannot be reduced simply by collecting more data. 
Figure~\ref{fig:backgr} shows the complete spectrum of IBD candidates with the fitted background contributions, while Fig.~\ref{fig:ffit} shows the background-subtracted $\rm E_{prompt}$\  spectrum along with the best fit.  
The pull parameters from the fit are summarized in Tab.~\ref{tab:pulls} together with the input values.
We have performed a frequentist study to determine the compatibility of the data and the no-oscillation hypothesis. 
Based on a $\Delta\chi^2$\  statistic, defined as the difference between the $\chi^2$\  at the best fit and at $\sin^2\!2\theta_{13}=0$, the data exclude the no-oscillation hypothesis at $97.4\,\%$\  ($2.0\,\sigma$).
A fit incorporating only the rate information yields $\sin^2\!2\theta_{13}=0.044\,\pm\,0.022\,\rm{(stat.)\,\pm\,0.056\,(syst.)}$. A simple ratio of observed to expected signal statistics yields $\rm R=0.978\,\pm\,0.011\,(stat.)\,\pm\,0.029\,(syst.)$ at the far site. 

The smaller best-fit value of  $\sin^2\!2\theta_{13}$\  by the rate-only analysis can be
explained by the $^{9}$Li background. The fit to the energy spectrum indicates a larger
$^{9}$Li background contamination than the original estimate, although it is consistent
within the systematic uncertainty.

\begin{table}[tbp]
\caption{Summary of pull parameters in the oscillation fit. The input values are determined by measurements, and the best-fit values are outcome of oscillation fit.}
\begin{tabularx}{\columnwidth}{lcc}
\hline
\hline
\textbf{Pull parameter} & \textbf{Initial} & \textbf{Best-fit }\\
& \textbf{value} &\textbf{value} \\
\hline
Cosmogenic isotope [day$^{-1}$] & $2.8\pm1.2$ & $3.9\pm0.6$\\
Fast neutrons [day$^{-1}$]      & $2.5\pm0.5$ & $2.6\pm0.4$\\
Energy scale       & $1.00\pm0.02$  & $0.99\pm0.01$\\
$\Delta m^2(10^{-3}\rm{eV}^2)$ & $2.32\pm0.12$ & $2.31\pm0.12$\\
\hline
\hline
\end{tabularx}
\label{tab:pulls}
\end{table}

%
%
In summary, due to the low level of backgrounds achieved in the Double Chooz detector, we have made  the first  measurement of    $\sin^2\!2\theta_{13}$ using the capture of  IBD neutrons on hydrogen. 
This technique enabled us to use a different data set with partially different systematic uncertainties than that used in the standard Gd analysis~\cite{DC2ndPub}. 
An analysis based on rate and spectral shape information yields   $\sin^2\!2\theta_{13}=0.097\,\pm\,0.034\,\rm{(stat.)\,\pm\,0.034\,(syst.)}$, 
which  is  in good agreement with the result of the Gd analysis $\sin^2\!2\theta_{13}=0.109\,\pm\,0.030\,\rm{(stat.)\,\pm\,0.025\,(syst.)}$~\cite{DC2ndPub}.
With increased statistics and a precise evaluation of the correlation of the systematic uncertainties, a combination of the two results is foreseen for the future. 

%
%
%

We thank the French electricity company EDF; the
European fund FEDER; the R\'egion de Champagne Ardenne;
the D\'epartement des Ardennes; and the Communaut\'e des Communes Ardennes Rives de Meuse. We
acknowledge the support of the CEA, CNRS/IN2P3, the
computer center CCIN2P3, and LabEx UnivEarthS in
France; the Ministry of Education, Culture, Sports, Science
and Technology of Japan (MEXT) and the Japan
Society for the Promotion of Science (JSPS); the Department
of Energy and the National Science Foundation of
the United States; the Ministerio de Ciencia e Innovaci\'on
(MICINN) of Spain; the Max Planck Gesellschaft,
and the Deutsche Forschungsgemeinschaft DFG (SBH
WI 2152), the Transregional Collaborative Research Center TR27, the 
excellence cluster ``Origin 
and Structure
of the Universe'', and the Maier-Leibnitz-Laboratorium
Garching in Germany; the Russian Academy of
Science, the Kurchatov Institute and RFBR (the Russian
Foundation for Basic Research); the Brazilian Ministry of
Science, Technology and Innovation (MCTI), the Financiadora
de Estudos e Projetos (FINEP), the Conselho
Nacional de Desenvolvimento Cient\'ifico e Tecnol\'ogico
(CNPq), the S\~ao Paulo Research Foundation (FAPESP),
and the Brazilian Network for High Energy Physics (RENAFAE)
in Brazil.


\begin{thebibliography}{10}

\bibitem{T2KTheta13}
K.~Abe {\it et al.}
\newblock {\em Phys. Rev. Lett.}, 107 041801 (2011).

\bibitem{MINOSTheta13}
P.~Adamson {\it et al.}
\newblock {\em Phys. Rev. Lett.}, 107, 181802 (2011).

\bibitem{DC1stPub}
Y.~Abe {\it et al.}
\newblock {\em Phys. Rev. Lett.}, 108, 131801 (2012).

\bibitem{DB1stPub}
F.P.~An {\it et al.}
\newblock {\em Phys. Rev. Lett.}, 108, 171803 (2012).

\bibitem{RENO1stPub}
J.K.~Ahn {\it et al.}
\newblock {\em Phys. Rev. Lett.}, 108, 191802 (2012).

\bibitem{DC2ndPub}
Y.~Abe {\it et al.}
\newblock {\em Phys. Rev. D}, 86, 052008 (2012).

\bibitem{DB2ndPub}
F.P.~An {\it et al.}
\newblock {\em Chinese Physics C}, 37, 011001 (2013).

\bibitem{scintillator}
C.~Aberle {\it et al.}
\newblock {\em JINST}, 7, P06008 (2012).

\bibitem{PMT1}
E.~Calvo {\it et al.}
\newblock {\em NIM A}, 621, 222 (2010).

\bibitem{PMT2}
C.~Bauer {\it et al.}
\newblock {\em JINST}, 6, P06008 (2011).

\bibitem{PMT3}
T.~Matsubara {\it et al.}
\newblock {\em NIM A}, 661, 16 (2012).

\bibitem{Bugey4}
Y.~Declais {\it et al.}
\newblock {\em Phys. Lett. B}, 338, 383 (1994).

\bibitem{MINOS:DM2}
P.~Adamson {\it et al.}
\newblock {\em Phys. Rev. Lett.}, 106, 181801 (2011).

\end{thebibliography}
\end{document}